\documentclass{jfm}

\usepackage{graphicx}
\usepackage{natbib}
\usepackage{color}

\ifCUPmtlplainloaded \else
  \checkfont{eurm10}
  \iffontfound
    \IfFileExists{upmath.sty}
      {\typeout{^^JFound AMS Euler Roman fonts on the system,
                   using the 'upmath' package.^^J}%
       \usepackage{upmath}}
      {\typeout{^^JFound AMS Euler Roman fonts on the system, but you
                   dont seem to have the}%
       \typeout{'upmath' package installed. JFM.cls can take advantage
                 of these fonts,^^Jif you use 'upmath' package.^^J}%
      }
  \else
  \fi
\fi

\ifCUPmtlplainloaded \else
  \checkfont{msam10}
  \iffontfound
    \IfFileExists{amssymb.sty}
      {\typeout{^^JFound AMS Symbol fonts on the system, using the
                'amssymb' package.^^J}%
       \usepackage{amssymb}%
         \let\leq=\leqslant
         \let\geq=\geqslant
      }{}
  \fi
\fi

\ifCUPmtlplainloaded \else
  \IfFileExists{amsbsy.sty}
    {\typeout{^^JFound the 'amsbsy' package on the system, using it.^^J}%
     \usepackage{amsbsy}}
    {}
\fi

\newcommand{\Ro}{\mathrm{Ro}}

\newcommand{\Fig}[1]{Fig.~\ref{#1}}
\newcommand{\fig}[1]{\Fig{#1}}
\newlength{\hfwidth}
\newlength{\hfwidthsingle}
\newcommand{\Eq}[1]{Eq. (\ref{#1})}
\newcommand{\eq}[1]{\Eq{#1}}

\newsavebox{\astrutbox}
\sbox{\astrutbox}{\rule[-5pt]{0pt}{20pt}}

\definecolor{brown}{rgb}{0.42,0.24,0.07}
\definecolor{darkgreen}{rgb}{0.0,0.6,0.00}

\title[Magneto-Elliptic-Rotational Instability]{On the connection between the magneto-elliptic and magneto-rotational instabilities}

\author[K. A. Mizerski and W. Lyra]%
{K\ls R\ls Z\ls Y\ls S\ls Z\ls T\ls O\ls F\ns A.\ns M\ls I\ls Z\ls E\ls R\ls S\ls K\ls I$^1$%
  \thanks{Email address for correspondence: kamiz@ippt.gov.pl},\ns
\and W\ls L\ls A\ls D\ls I\ls M\ls I\ls R\ns L\ls Y\ls R\ls A$^{2,3,4}$}

\affiliation{$^1$Department of Mechanics and Physics of Fluids, Institute of Fundamental and Technological Research,
Polish Academy of Sciences, Pawinskiego 5B, 02-106, Poland\\[\affilskip]
$^2$Department of Astrophysics, American Museum of Natural History, 79th Street at Central Park West, New York NY 10024-5192, USA\\[\affilskip]
$^3$Jet Propulsion Laboratory, 4800 Oak Grove Drive, Pasadena CA 91109, USA\\[\affilskip]
$^4$NASA Carl Sagan Fellow}

\date{11 July 2011; Accepted 15 February 2012.}
\begin{document}

\maketitle

\begin{abstract}
It has been recently suggested that the magneto-rotational 
instability (MRI) is a limiting case of the magneto-elliptic 
instability (MEI). This limit 
is obtained for horizontal modes in the presence of rotation and 
an external vertical magnetic field, when the aspect ratio of the 
elliptic streamlines tends to infinite. In this 
paper we unveil the link between these previously unconnected mechanisms, 
explaining both the MEI and the MRI as different manifestations of the same 
{\it Magneto-Elliptic-Rotational Instability} (MERI). The 
growth rates are found and the influence of the magnetic and rotational 
effects is explained, in particular the effect of the magnetic field 
on the range of negative Rossby numbers at which the horizontal 
instability is excited. Furthermore, we show how the horizontal 
rotational MEI in the rotating shear flow limit links to the MRI by 
the use of the local shearing box model, typically used in the 
study of accretion discs. In such limit the growth rates of the two 
instability types coincide for any power-type background angular 
velocity radial profile with negative exponent corresponding to 
the value of the Rossby number of the rotating shear flow. The MRI 
requirement for instability is that the background angular velocity 
profile is a decreasing function of the distance from the centre 
of the disk which corresponds to the horizontal rotational MEI requirement 
of negative Rossby numbers. Finally a physical interpretation of 
the horizontal instability, based on a balance between the strain, the Lorentz force
and the Coriolis force is given.
\end{abstract}


\section{Introduction}

The stability of vortices is a problem of paramount importance in fluid 
mechanics. Considering 
that turbulence consists of vorticity blobs, vortices are the fundamental 
unit of turbulent flow. Unveiling the mechanism that renders them unstable 
should therefore provide vital insights into the nature of turbulence 
itself. Phenomenologically, turbulence can be described as a series of 
bifurcations, starting with a primary instability that converts shear 
into vorticity, creating vortices. This is followed by another 
bifurcation, a secondary instability, to break these vortices into lesser 
vortical structures. These in turn shall experience a sequence of ``inertial 
instabilities'', leading to a cascade. Though the Kelvin-Helmholtz instability 
and the Rayleigh-Taylor instability are well established as examples 
of primary instabilities, the highly successful theory of the turbulent 
cascade put forth by Kolmogorov (1941) rested 
on a heuristic picture of secondary instability, established 
by early experiments (e.g., Taylor 1923). It was not until the 
works of Pierrehumbert (1986) and Bayly (1986) that the elliptic 
instability was introduced as a mechanism for the secondary 
instability. A fluid in rigid rotation supports a spectrum of 
stable inertial waves, the simplest case being circularly 
polarized transverse plane waves oscillating at twice the 
frequency of the base flow. Strain is introduced when the streamlines 
pass from circular to elliptical, and some modes find resonance with 
the strain field, leading to de-stabilization. The nonlinear evolution and saturation of the elliptic instability was studied numerically by Schaeffer \& Le Diz$\grave{\textrm{e}}$s (2010) and experimentally by Eloy \textit{et al.} (2003). Herreman \textit{et al.} (2010) generalized the linear and weakly nonlinear theories for the elliptic instability (Waleffe 1989) to the magnetohydrodynamics case and conducted experiments to explain some aspects
of the nonlinear transition process.

The seminal work of Pierrehumbert (1986) and Bayly (1986) was followed 
by studies that considered the incorporation of more physics into the problem. 
Miyakazi (1993) included background rotation, unveiling a strong 
two-dimensional mode, for which the wavevector and the vortex axis 
are parallel. For incompressible flow such modes are solenoidal, so 
the motion itself is horizontal to the vortex axis. This horizontal 
instability is thus the two-dimensional mode of what could be called 
an elliptic-rotational instability (ERI). This instability is not of 
resonant nature, but centrifugal, appearing as exponential growth of 
epicyclic disturbances. This behaviour invites a connection with the 
Rayleigh (centrifugal) instability, and indeed the mechanism is similar (Lesur 
\& Papaloiozou 2009), suggesting that the Rayleigh instability is 
a limit of the ERI.

If the elliptic instability is such a fundamental process for 
understanding hydrodynamical turbulence, one can expect that the 
magneto-elliptic instability (MEI) should likewise provide a similar 
framework when it comes to MHD turbulence. In the geophysical context, 
Kerswell (1993) studied the influence of a toroidal magnetic field and 
stratification, both axial and radial, on the stability of elliptic flow 
in uniform rotation. In his configuration, axial stratification and the 
magnetic field were invariably stabilizing, though radial stratification 
enhanced the elliptical instability. Lebovitz \& Zweibel (2004) 
studied the problem of stability of axially magnetized elliptic streamlines, 
finding that the MEI has new families of destabilizing resonances, 
corresponding to Alfv\'en waves. Just as 
Miyakazi (1993) extended the work of Bayly (1986) to include background 
rotation, the work of Lebovitz \& Zwebeil (2004) was similarly extended 
by Mizerski \& Bajer (2009), who analysed the joint effect of the
magnetic and rotational effects on the stability of the Euler flow with 
elliptical streamlines, by the use of numerical and asymptotic analytical 
methods in the limit of small ellipticity of the flow. Their comprehensive 
study, based on the analytic technique developed by Lebovitz \& Zweibel (2004) 
included both the modes excited via a resonance mechanism and the 
horizontal modes, that propagate along the magnetic field 
lines. The case of the horizontal instability was then further investigated 
in the non-magnetic case by Lesur \& Papaloizou (2009) and 
in the magnetized case by Bajer \& Mizerski (2011). However, the problem 
of the MEI in the presence of rotation is conceptually difficult and there 
are aspects which still require clarification and further insight is 
necessary. 

One such insight may come from astrophysics, which also provides 
the motivation for the present study. When modeling the problem 
of magnetized vortices in protoplanetary disks, Lyra \& Klahr (2011) found that 
the horizontal magneto-elliptic mode, when taken to the limit of infinite 
ellipticity, yields the same growth rates and range of unstable wavenumbers 
as the well known magneto-rotational instability (MRI, Velikhov 1959, 
Chandrasekhar 1960). The MRI is an instability of magnetized Euler flow 
in circular differential rotation, which renders the motion unstable when the 
angular velocity decreases outwards. Since Keplerian flow is an Euler 
flow in circular differential rotation, and magnetic fields are 
ubiquitous in astrophysics, the MRI is the best candidate to generate 
turbulence in accretion disks (Balbus \& Hawley 1991, 1998, Armitage 1998, 
Hawley 2000, Fromang \& Nelson 2006, Lyra et al. 2008, Flock et al. 2011). 

The result of Lyra \& Klahr (2011) tantalizingly suggests that the MRI is 
a limiting case of the MEI. In this paper we explore this interesting 
prospect in more detail, unveiling the connection between these two previously 
unconnected instabilities. In a way, the present study is an attempt at 
unification, explaining both instabilities as different manifestations of 
the same {\it Magneto-Elliptic-Rotational Instability} (MERI). 

The paper is structured as follows. In the next section we briefly review 
the horizontal rotational MEI presenting a somewhat different approach from Bajer 
\& Mizerski (2011) and show that the external magnetic field has a profound 
effect on the range of values of the Rossby number for which the horizontal 
rotational MEI is excited. In Sect. 3 we further develop the idea presented in 
Lyra \& Klahr (2011) and demonstrate, by the use of the so-called local 
shearing box model (cf. Hawley et al. 1995) how the horizontal rotational MEI, in the 
limit of rotating shear flow can be related to the MRI. Next we present a 
physical interpretation of the horizontal instability based 
on a balance between strain generated by the basic flow and the Coriolis and Lorentz
forces. Conclusions are presented 
in Sect. 5. 

\section{The calculation of growth rates of the horizontal rotational MEI}

\begin{figure}
  \begin{center}
    \resizebox{.6\textwidth}{!}{\includegraphics{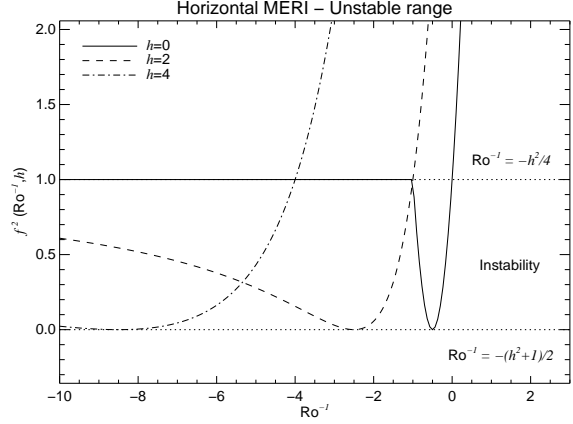}}
  \end{center}
  \caption[]{The square of the function $\psi_+$ as in \eq{eq:HIfg}, that defines the unstable 
      horizontal modes of the magneto-elliptic-rotational instability. Instability 
      exists when $|\psi_+| < \varepsilon$, where $0 \leq \varepsilon \leq 1$  is a measure of eccentricity. 
      The instability exists only for negative Rossby numbers, i.e., anticyclonic 
      rotation. The unstable range of wavenumbers is $\Ro^{-1} < -h^2/4$ and unbounded.}
  \label{fig:hmodes}
\end{figure}

Let us consider the Euler flow with elliptic streamlines,
\begin{equation}
\mathbf{u}_0=\gamma\left[-\left(1+\varepsilon\right)y,\,\left(1-\varepsilon\right)x\right]\label{eq:BF}
\end{equation}
\noindent rotating with the angular velocity $\mathbf{\Omega}=\Omega\hat{\mathbf{e}}_{z}$, where $2\gamma\hat{\mathbf{e}}_{z}$ is its uniform vorticity and $0<\varepsilon<1$ is the strain. We now briefly review the stability properties of the flow (\ref{eq:BF}) in the presence of uniform vertical field $\mathbf{B_0}=B_0\hat{\mathbf{e}}_{z}$ with respect to
the horizontal modes, which are defined by the fact that they propagate only
in the vertical direction, i.e. the wave vector of the perturbation
has only the `$z$'-component. The solenoidal constraints for the
velocity and magnetic fields then imply that the perturbations have
only horizontal components. Thus the velocity and magnetic field perturbations have the form
\begin{equation}
\mathbf{v}=\hat{\mathbf{v}}\mathrm{e}^{\mathrm{i}kz}\qquad \textrm{and}\qquad \mathbf{b}=\hat{\mathbf{b}}\mathrm{e}^{\mathrm{i}kz}\end{equation}
\noindent respectively. The equations to solve 
are the Euler and induction equations
\begin{equation}
\partial_t v_i + v_j \partial v_i = -\frac{1}{\rho}\partial_i P - 2 O_{ij} v_j + \frac{1}{\rho\mu_0}B_k\partial_k B_i,\label{eq:euler}\end{equation}
\begin{equation}
\partial_t B_i + v_j\partial_j B_i = B_j\partial_j B_i,\label{eq:induction}\end{equation}
\noindent where the magnetic pressure and centrifugal potential were 
incorporated in the pressure $P$ and $O_{ik} = \varepsilon_{ijk} \varOmega_j$. We apply the scaling $b_i \rightarrow b_i/\sqrt{\mu_0\rho}$ so that, 
in principle, we are solving for the Alfv\'en velocity and after linearisation the evolution equations for the horizontal rotational MEI modes yield
\begin{equation}
  \left[\begin{array}{c}
      \dot{\hat{v}}_{x}\\
      \dot{\hat{v}}_{y}\\
      \dot{\hat{b}}_{x}\\
      \dot{\hat{b}}_{y}
    \end{array}\right]=\left[\begin{array}{cccc}
      0 & 1+\varepsilon+2\Ro^{-1} & ih & 0\\
      -\left(1-\varepsilon\right)-2\Ro^{-1} & 0 & 0 & ih\\
      ih & 0 & 0 & -\left(1+\varepsilon\right)\\
      0 & ih & 1-\varepsilon & 0
    \end{array}\right]\left[\begin{array}{c}
      \hat{v}_{x}\\
      \hat{v}_{y}\\
      \hat{b}_{x}\\
      \hat{b}_{y}
    \end{array}\right]\label{eq:HI_equations}
\end{equation}
\noindent where $\tau=t/\gamma$ and the upper dot denotes a derivative with respect to $\tau$, $\Ro=\gamma/\Omega$ is the Rossby
number and $h=kB_{0}/\gamma\sqrt{\mu_{0}\rho}$. The eigenvalues of
the matrix in (\ref{eq:HI_equations}) which we denote by $\sigma_{j}$,
$j=1,2,3,4$ are the complex growth rates of the perturbations. They are
\begin{subequations}\label{eq:HIsigma}
  \begin{equation}
    \sigma_{1}\left/\gamma\right.=\sqrt{\varepsilon^{2}-\psi_{+}^{2}}, \qquad \sigma_{2}\left/\gamma\right.=-\sqrt{\varepsilon^{2}-\psi_{+}^{2}},\end{equation}
  \begin{equation}
    \sigma_{3}\left/\gamma\right.=\sqrt{\varepsilon^{2}-\psi_{-}^{2}}, \qquad \sigma_{4}\left/\gamma\right.=-\sqrt{\varepsilon^{2}-\psi_{-}^{2}},\end{equation}
\end{subequations}
\noindent where
\begin{equation}
  \psi_{\pm} = \Ro^{-1}\pm\sqrt{\left(\Ro^{-1}+1\right)^{2}+h^{2}}.\label{eq:HIfg}
\end{equation}
\noindent Note that for all $h\geq0$ and $\Ro\in\mathbb{R}$
\begin{equation}
  \left|\psi_{-}\right|\geq1\geq\varepsilon,\label{eq:cond_g}
\end{equation}
\noindent which means that $\sigma_3$ and $\sigma_4$ are imaginary, thus produce only oscillations. The basic state is unstable with respect
to perturbations propagating along the `$z$' axis if and only if there exist $\Ro$ and $h$ such that
\begin{equation}
\left|\psi_{+}\right|<1\qquad\textrm{and}\qquad\varepsilon\geq\left|\psi_{+}\right|.\label{eq:cond_destab}
\end{equation}
\noindent The function $\psi_{+}\left(\Ro^{-1};h\right)$, plotted in 
\fig{fig:hmodes}, possesses the following properties

\begin{subequations}\label{eq:prop_chi}
  \begin{eqnarray}
    \psi_{+}\geq 1\,\,;&& \textrm{for} \,\,\,\,\Ro^{-1}\geq-\frac{h^{2}}{4}\\
    \psi_{+}=1    \,\,;&& \textrm{for} \,\,\,\,\Ro^{-1}=-h^2/4      \\
    0\leq\left|\psi_{+}\right|<1\,\,; && \textrm{for} \,\,\,\,\Ro^{-1}<-\frac{h^{2}}{4}\\
    \psi_{+}=0    \,\,;&& \textrm{for} \,\,\,\,\Ro^{-1}=-(h^2+1)/2
\end{eqnarray}
\end{subequations}

\noindent and
\begin{eqnarray}
\lim_{\Ro^{-1}\rightarrow-\infty}&&\psi_{+}\left(\Ro^{-1};h\right)=\\\nonumber
\lim_{\Ro^{-1}\rightarrow-\infty}&&\Ro^{-1}\left\{ 1-\left[1+\Ro+\mathcal{O}\left(\Ro^{2}\right)\right]\right\}= -1.\label{eq:limes_f}
\end{eqnarray}
\noindent Therefore the instability is only possible if $\Ro^{-1}<-h^{2}/4$
since only then, for a certain ellipticity $\varepsilon\geq\left|\psi_{+}\right|$
the growth rate $\sigma_{1}$ can be real and positive.\textbf{\LARGE{}
}Moreover, equation (\ref{eq:limes_f}), in particular, means that
as $\Ro^{-1}$ tends to $-\infty$, for some $\Ro^{-1}<-(h^{2}+1)/2$
the function $\psi_{+}^{2}(\Ro^{-1};h)$ must have an inflection point. The
limit $h\rightarrow0$ is singular in the sense that the two curves
$\psi_{+}^{2}(\Ro^{-1};h)$ and $\psi_{-}^{2}(\Ro^{-1};h)$ merge into a parabola
(and a straight line parallel to the $\Ro^{-1}$ axis of constant value
1), touching each other at the inflection point of the function $\psi_{+}^{2}(\Ro^{-1};h)$
at $\Ro^{-1}=-1$, $\psi_{+}^{2}=\psi_{-}^{2}=1$. This means that for $h=0$ the
horizontal instability is possible only for a bounded set of values
$\Ro^{-1}\in(-1,0)$. The situation significantly changes immediately
after switching on even a very weak magnetic field since then the
domain in which the horizontal instability is possible becomes unbounded,
i.e. $\Ro^{-1}\in(-\infty,-h^{2}/4)$. As the instability exists only 
for negative Rossby numbers, only anticyclonic vortices are unstable.

As Lyra \& Klahr (2011) have pointed out, there is instability in the 
limit of a vanishing vortex in a shearing box (cf. Hawley et al. 1995). In that case, 
the Rossby number remains finite owing to the vorticity of the Keplerian shear, and the 
case is equivalent to a Kida vortex (Kida 1981) of infinite aspect ratio
\begin{equation}
  \varepsilon=1\qquad\textrm{and}\qquad \Ro=-\frac{3}{4}\label{eq:SFKLimit}
\end{equation}
In the limit $\varepsilon=1$ only the horizontal instability persists (the purely horizontal modes are the most unstable ones and the modes destabilised via resonances are suppressed; cf. Mizerski \& Bajer 2009, 2011), 
for which the dispersion relation is (cf. equation (\ref{eq:HIsigma}a))
\begin{equation}
  \frac{\sigma^{2}}{\gamma^{2}}=\varepsilon^{2}-\left[\Ro^{-1}+\sqrt{\left(\Ro^{-1}+1\right)^{2}+\frac{k^{2}v_{A}^{2}}{\gamma^{2}}}\right]^{2},\label{eq:grate_1}
\end{equation}
\noindent where $v_{A}=B_0/\sqrt{\mu_0\rho}$ is the Alfv$\acute{\textrm{e}}$n speed. For $\Ro<0$, also substituting $\gamma=\Ro\,\Omega$ and $q=kv_{A}/\Omega$, we obtain
\begin{equation}
  \frac{\sigma^{2}}{\gamma^{2}}=\varepsilon^{2}-\Ro^{-2}\left[\sqrt{\left(\Ro+1\right)^{2}+q^{2}}-1\right]^{2},\label{eq:grate 2}
\end{equation}
\noindent and applying the limits defined by (\ref{eq:SFKLimit})
\begin{equation}
  \frac{\sigma^{2}}{\gamma^{2}}=\frac{8}{9}\left(\sqrt{16q^{2}+1}-2q^{2}-1\right).\label{eq:grate_SFKLimit}
\end{equation}
\noindent Thus the growth 
rate is real and positive for $0<q<\sqrt{3}$, and the maximal 
value $\sigma=\left|\gamma\right|=(3/4)\left|\Omega\right|$
is achieved for $q=\sqrt{15}/4\approx0.9682$. This 
is precisely the same dispersion relation obtained for the MRI by 
Balbus \& Hawley (1991, see also Hawley \& Balbus 1991). 

\section{Why do we see the MRI in a rotating shear flow?}

\begin{figure}
  \begin{center}
    \resizebox{.3\textwidth}{!}{\includegraphics{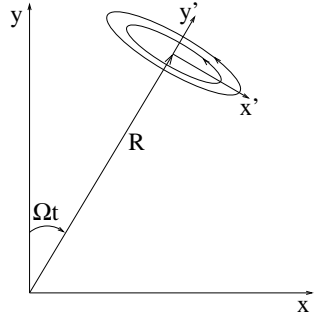}}
  \end{center}
  \caption[]{A schematic picture of the rotating elliptic flow.}
  \label{fig:rotellipse}
\end{figure}

The MRI was originally formulated for axisymmetric flows with circular
streamlines and the condition for MRI to operate is that the angular
velocity decreases with radius. However, as it turns out, the MRI
also seems to operate in non-axisymmetric flows and appears
as a limiting case of the MEI in the presence of rotation, in the limit when the latter becomes a rotating shear
flow (Lyra \& Klahr 2011). Let us now assume that the Euler flow with elliptic streamlines given in (\ref{eq:BF}),
rotates with the angular velocity $\mathbf{\Omega}=\Omega\hat{\mathbf{e}}_{z}$
about an axis parallel the `$z$' axis and located at distance $R$
from the centre of the elliptical vortex. Let us also consider the
inertial (non-rotating) system of reference ($x'$,$y'$,$z'$), with
the origin $x'=0$, $y'=0$ at the axis of rotation (see \fig{fig:rotellipse}).
In this frame of reference the vector of position of the centre of
the elliptical vortex is
\begin{equation}
  \mathbf{R}'=R\left[-\sin\Omega t,\,\cos\Omega t\right],\label{eq:R}
\end{equation}
\noindent where the prime indicates representation in the non-rotating frame and
\begin{equation}
  \frac{\mathrm{d}\mathbf{R}'}{\mathrm{d}t}=\mathbf{\Omega}\times\mathbf{R}',\label{eq:R_eq}
\end{equation}
\noindent where the time derivative is taken within the non-rotating frame, i.e. holding the unit vectors $\hat{\mathbf{e}}'_{x}$ and $\hat{\mathbf{e}}'_{y}$ constant. It was chosen that
\begin{equation}
  \mathbf{R}'\left(t=0\right)=R\hat{\mathbf{e}}'_{y}.\label{eq:R_ic}
\end{equation}
\noindent Using the transformations
\begin{equation}
  \left[\begin{array}{c}
      x\\
      y
    \end{array}\right]=\left[\begin{array}{cc}
      \cos\Omega t & \sin\Omega t\\
      -\sin\Omega t & \cos\Omega t
    \end{array}\right]\left[\begin{array}{c}
      x'\\
      y'
    \end{array}\right]-\left[\begin{array}{c}
      0\\
      R
    \end{array}\right],\qquad\begin{array}{c}
  \hat{\mathbf{e}}_{x}=\cos\Omega t\hat{\mathbf{e}}'_{x}+\sin\Omega t\hat{\mathbf{e}}'_{y}\\
  \hat{\mathbf{e}}_{y}=-\sin\Omega t\hat{\mathbf{e}}'_{x}+\cos\Omega t\hat{\mathbf{e}}'_{y}
  \end{array}\label{eq:trans1}
\end{equation}
\noindent and
\begin{equation}
  \mathbf{u}_{0}^{\prime}=\mathbf{u}_0+\mathbf{\Omega}\times\mathbf{r}'+\frac{\mathrm{d}\mathbf{R}'}{\mathrm{d}t}-\mathbf{\Omega}\times\mathbf{R}'=\mathbf{u}+\mathbf{\Omega}\times\mathbf{r}'\label{eq:trans2}
\end{equation}
\noindent where $\mathbf{r}'=\left[x',y',z'\right]^{T}$, we obtain the
following expression for the basic elliptic velocity field in the inertial frame
\begin{equation}
  u_{0x}^{\prime}=-\gamma\left[y'+\varepsilon\left(y'\cos2\Omega t-x'\sin2\Omega t\right)\right]+\gamma\left(1+\varepsilon\right)R\cos\Omega t-\Omega y',\label{eq:BF_NRF_u}
\end{equation}
\begin{equation}
  u_{0y}^{\prime}=\gamma\left[x'-\varepsilon\left(x'\cos2\Omega t+y'\sin2\Omega t\right)\right]+\gamma\left(1+\varepsilon\right)R\sin\Omega t+\Omega x'.\label{eq:BF_NRF_v}
\end{equation}
\noindent Using the cylindrical coordinates ($r'$,$\varphi'$,$z$)
we may write the angular velocity of the flow in the inertial frame
as
\begin{equation}
  \frac{u_{0\varphi}^{\prime}}{r'}=\frac{u_{0y}^{\prime}x'-u_{0x}^{\prime}y'}{r^{\prime 2}}=\Omega\left\{ 1+\Ro\left[1-\varepsilon\cos\left(2\varphi'-2\Omega t\right)\right]-\Ro\left(1+\varepsilon\right)\frac{R}{r'}\sin\left(\varphi'-\Omega t\right)\right\} ,\label{eq:Ang_Vel}
\end{equation}
\noindent which, clearly, is a function of $r'$ (as long as $R\neq0$)
and $\varphi'$ (in the above $\Ro=\gamma/\Omega$ is the Rossby number).
Lyra \& Klahr (2011) have shown numerically that in the rotating shear flow limit defined
in (\ref{eq:SFKLimit}) the rotational MEI growth rate (i.e. the horizontal 
instability growth rate) profile with the parameter $q$ indeed 
matches the MRI profile. In such limit the expression (\ref{eq:Ang_Vel}) yields
\begin{equation}
  \frac{u^{\prime}_{0\varphi}}{r'}=\Omega\left\{ \frac{1}{4}+\frac{3}{4}\cos\left(2\varphi'-2\Omega t\right)+\frac{3}{2}\frac{R}{r'}\sin\left(\varphi'-\Omega t\right)\right\} ,\label{eq:Ang_Vel_SFKLimit}
\end{equation}
\noindent therefore the angular velocity in the inertial frame, in
the rotating shear flow limit, is depended on the distance from the
origin as long as $R\neq0$. Moreover, the flow in the inertial frame
is not axially symmetric, but nevertheless the resemblance to the MRI
is observed. Furthermore the linear stability analysis
which leads to the expressions for the growth rate (\ref{eq:grate_1})
and thus (\ref{eq:grate 2}) is independent of the value of $R$ which
means that even for $R=0$, i.e. when the angular velocity is independent
of $r'$ we obtain the MRI-like results in the rotating shear flow
limit of the rotational MEI. Since the Rossby number must be negative
for the instability to settle in, it seems that it is crucial that
the MRI requirement for the angular velocity to be a decreasing
function of $r'$ must be met in the neighborhood of the centre of
the rotating elliptic vortex, i.e. for $x=0$, $y=0$. In principle
for a given value of the Rossby number the angular velocity (\ref{eq:Ang_Vel_SFKLimit})
is a decreasing function of $r'$ only in a half-plane and an increasing
function of $r'$ in the other half-plane, depending on the signs
of $\Ro$ and $\sin\left(\varphi'-\Omega t\right)$. However, as said,
the instability occurs only for $\Ro<0$, in which case the angular
velocity decreases with $r'$ for $\Omega t<\varphi'<\pi+\Omega t$,
and the steepest decrease is observed at $\varphi'=\pi/2+\Omega t$,
which is the azimuthal coordinate of the centre of the elliptic vortex.

\noindent The explanation is based on the local shearing sheet model
for accretion disks, since in this model the shearing flow is simply
a local representation of the background rotation of the axisymmetric
vortex with angular velocity decreasing with $r'$. In other words
for an axisymmetric vortex with the angular velocity depended on
$r'$ as $\Omega(r')\sim1/r^{\prime s}$ and thus gravity as $g\sim r^{\prime n}$,
with $n=-2s+1$, in the vicinity of $r'=R$ we get
\begin{equation}
  \Omega(r)\sim\frac{1}{\left(R+r\right)^{s}}\approx\frac{1}{R^{s}}-sr\frac{1}{R^{s+1}},\label{eq:AV_exp_KD}
\end{equation}
\noindent and from (\ref{eq:Ang_Vel}), in the vicinity of $\mathbf{r}'=\mathbf{R}'$,
i.e. for $r'=R+r$ and $\varphi'=\pi/2+\Omega t+\varphi$, we obtain
\begin{equation}
  \frac{u_{0\varphi}^{\prime}}{r'}=\Omega\left\{ 1+\Ro+\Ro\cos\left(2\varphi\right)-2\Ro\cos\varphi+2\Ro\frac{r}{R}\cos\varphi\right\} .\label{eq:AV_exp_1}
\end{equation}
\noindent According to the shearing sheet approximation we now substitute
in (\ref{eq:AV_exp_1}) for $\Omega=1/R^{s}$ and since $\cos\varphi\approx1+\mathrm{O}(\varphi^{2})$
the expression (\ref{eq:AV_exp_1}) up to order one in the distance
from the point $\mathbf{r}'=\mathbf{R}'$ (thus in the region of validity
of the shearing sheet approximation) takes the form
\begin{equation}
  \frac{u_{0\varphi}^{\prime}}{r'}\approx\frac{1}{R^{s}}+2\Ro\frac{r}{R^{s+1}},\label{eq:AV_exp_2}
\end{equation}
\noindent which agrees with (\ref{eq:AV_exp_KD}) for
\begin{equation}
  \Ro=-\frac{s}{2}=\frac{n-1}{4}.\label{eq:Ro_SSA}
\end{equation}
\noindent The Kida (1981) solution for a stationary elliptic vortex
embedded in a shear flow yields
\begin{equation}
  \Ro=-\frac{s}{2}\frac{2}{\left(1+\varepsilon-\sqrt{1-\varepsilon^{2}}\right)},\label{eq:Ro_Kida}
\end{equation}
\noindent which in the shear flow limit $\varepsilon=1$ agrees with
(\ref{eq:Ro_SSA}). The horizontal rotational MEI in the presence
of a magnetic field operates for all negative Rossby numbers and according
to (\ref{eq:Ro_SSA}) this corresponds to $s>0$, i.e. the angular
velocity $\Omega(r')$ decreasing with $r'$, which is the requirement
for the MRI. It is now clear, that since the growth rate
of the horizontal instability in a rotating shear flow (\ref{eq:grate_1})
does not depend on position, and at the centre of the elliptical vortex
$\mathbf{r}'=\mathbf{R}'$ the rotating shear flow models locally the
flow of an axisymmetric disk with angular velocity decreasing with
$r'$, the growth rate must correspond to the growth rate of the
MRI taken at $\mathbf{r}'=\mathbf{R}'$, since in the short-wavelength
limit analysed by Balbus \& Hawley (1991) the MRI criterion
is local. Nevertheless it remains an interesting feature of the rotating
shear flow limit that despite the fact that the flow is not axially
symmetric and for $R=0$ can even be independent of $r'$ the stability
characteristics are the same as of the classic axially symmetric,
short-wavelength MRI.

It is also important to realise that, in general, the horizontal instability
persists even without the presence of the magnetic field. The MRI
operates only if the magnetic field is present. Furthermore,
as already mentioned, even the $r'$-dependence of the angular velocity
is not necessary for the development of the horizontal instability,
since if $R=0$ the expression (\ref{eq:Ang_Vel}) becomes independent
of $r'$ and the growth rate (\ref{eq:grate_1}) is not affected.
Thus, in principle, it is the lack of axial symmetry that is crucial
for the development of the horizontal rotational MEI. Nevertheless
there exists a common physical ground for both the instability types,
i.e. horizontal rotational MEI and MRI, since the rotating shear flow limit of
the rotational MEI, even though non-axially symmetric, exhibits the
properties of the MRI. The common feature of both 
instability types is the shear which is present in the
elliptic flows as well as in axisymmetric flows with angular velocity
dependent on radius and this feature accompanied by the rotational
effects (local in MRI and global in rotational MEI) is behind the physical interpretation
of the common limit for both instabilities discussed in here. 
This point is elaborated in the next Section.

We have, therefore, demonstrated the direct link between the megneto-elliptic and magneto-rotational instabilities and in the non-magnetic case between the horizontal elliptic and centrifugal instabilities and the physical conditions under which previously unrelated instability types can be linked were explained in detail. The dispersion relations for horizontal MEI and MRI, likewise for horizontal EI and the centrifugal instability are the same in the limit $\epsilon=1$, which suggests, that the physics of the horizontal magneto-elliptic instability is similar to the physics of the MRI.

Let us make a final comment on the non-magnetic case. As mentioned
the horizontal instability is present even if the magnetic field is
not. The axisymmetric flows in the non-magnetic case are only subject
to the centrifugal instability for which the Rayleigh's criterion
says that instability can occur if the angular momentum is a decreasing
function of $r'$ in some region. According to (\ref{eq:AV_exp_KD}),
(\ref{eq:AV_exp_2}) and (\ref{eq:Ro_SSA}) the angular momentum of
an axisymmetric vortex decreases with $r'$ if the Rossby number for
the rotating shear flow satisfies $\Ro<-1$. This is an exact instability
condition for the horizontal modes in the non-magnetic case (cf. Mizerski
\& Bajer 2009) for a rotating shear flow, i.e. for $\varepsilon=1$, since
in such case the growth rate of the horizontal modes is (cf. (\ref{eq:grate_1}) and (\ref{eq:Ro_SSA}))
\begin{equation}
  \sigma^{2}=2\left(s-2\right)\Omega^{2}\left(R\right).\label{eq:grate_HI_CI}
\end{equation}
\noindent On the other hand, the classic Rayleigh's criterion, formulated
for systems with impermeable boundaries at $r'=R_{1}$ and $r'=R_{2},$ for
horizontal modes and $\Omega(r')\sim1/r^{\prime s}$ yields
\begin{equation}
  \sigma^{2}=-2\frac{s-2}{s-1}\frac{R_{2}^{2-2s}-R_{1}^{2-2s}}{R_{2}^{2}-R_{1}^{2}}.\label{eq:grate_CI}
\end{equation}
\noindent If the Rayleigh's criterion is made local, by assuming a
narrow gap limit, i.e. $R_{2}=R_{1}+\delta r$ and $\delta r/R_{1}\ll1$
the above expression (\ref{eq:grate_CI}) reduces to
\begin{equation}
  \sigma^{2}=2\left(s-2\right)\Omega^{2}\left(R_{1}\right),\label{eq:grate_CI_local}
\end{equation}
\noindent which agrees with (\ref{eq:grate_HI_CI}).

It seems in place to briefly comment on the case when the applied field in a disc is azimuthal, since the numerical and experimental works of Hollerbach \& R$\ddot{\textrm{u}}$diger (2005) and Stefani \textit{et al.} (2006) suggest a great importance of the azimuthal field for the reduction of the instability thereshold. For the rotating shear flow $\mathbf{u}_0=-2\gamma y\hat{\mathbf{e}}_{x}$ modeling locally the flow in a disc one could take $\mathbf{B_0}=(B_0+J_0y)\hat{\mathbf{e}}_{x}$. Such stability problem with perturbations in the form of horizontal modes takes the form
\begin{equation}
  \frac{\sigma}{\gamma}\left[\begin{array}{c}
      \hat{v}_{x}\\
      \hat{v}_{y}\\
      \hat{b}_{x}
    \end{array}\right]=\left[\begin{array}{cccc}
      0 & 2\left(1+\Ro^{-1}\right) & 0\\
      -2\Ro^{-1} & 0 & -\Gamma\\
      0 & -\Gamma & 0\\
    \end{array}\right]\left[\begin{array}{c}
      \hat{v}_{x}\\
      \hat{v}_{y}\\
      \hat{b}_{x}
    \end{array}\right],\label{eq:equations_azimuthal_field}
\end{equation}
with $\hat{b}_{y}=0$ and $\Gamma=J_0/(\gamma\sqrt{\mu_0\rho})$. Solving for the eigenvalue one obtaines
\begin{equation}
\frac{\sigma^2}{\gamma^2}=-4Ro^{-1}\left(1+Ro^{-1}\right)+\Gamma^2,\label{grate_az_field}
\end{equation}
\noindent which leads to a conclusion that the instability is possible if either $Ro<-2/(1+\sqrt{1+\Gamma^2})$ or $Ro>2/(\sqrt{1+\Gamma^2}-1)$ is satisfied. Thus a new unstable branch appears for large and positive Rossby numbers, which has the property, that for weak field gradients, $\Gamma\ll 1$ only strongly shearing discs are affected, $Ro=-s/2\gg 1$. In principle the correspondence between the stability characteristics of the rotating shear flow with horizontal field and the local stability characteristics of a disc with azimuthal field should exist, however, the horizontal modes anaysed here are independent of $y$, thus axisymmetric in the global sense and as argued by Terquem \& Papaloizou (1996) the most unstable modes in the case of azimuthal fields are the non-axisymmetric ones. Therefore the dynamics of discs with azimuthal basic fields should be dominated by the growth of non-axisymmetric modes not analysed in here.

\section{Physical interpretation of the horizontal rotational MEI}

The essence of the rotational MEI is the departure from axial symmetry, which imposes strain on the perturbations. The typical model based on consideration of radial perturbations which preserve angular momentum, useful for physical interpretation of the MRI and the centrifugal instability is no longer applicable when the vortex is elliptical. A similarly simple physical model of the rotational MEI would have to take into account not only the elliptical shape of the vortex but also the action of the Coriolis force, which is essential for the instability mechanism. We present a somewhat simpler approach and provide the physical explanation of the horizontal rotational MEI based on the balance of forces acting on a general horizontal perturbation, highlighting the importance of strain. 
Let us write down the equations governing the evolution of the horizontal
modes, for which the wave vector is vertical, $\mathbf{k}=k\hat{e}_{z}$,
and thus, because of the divergence free constraint, the perturbations
posses only the horizontal components. The Navier-Stokes equation
in the rotating frame and in the non-magnetic case is
\begin{equation}
  \frac{\partial\mathbf{v}}{\partial t}+\left(\mathbf{v}\cdot\nabla\right)\mathbf{u}_0=-2\mathbf{\Omega}\times\mathbf{v},\label{eq:pert_eq_gen}
\end{equation}
\noindent where $\mathbf{u}_0$ is the basic elliptic flow defined
in (\ref{eq:BF}) and $\mathbf{v}$ is the horizontal
perturbation in the rotating frame. Note that the term describing
advection of the perturbation by the elliptic flow, $(\mathbf{u}_0\cdot\nabla)\mathbf{v}$
and the pressure terms vanish in the case of purely horizontal perturbations
and the only terms that remain are the terms describing the stretching
of the perturbation by the basic flow and the Coriolis force. Introducing
$\mathbf{v}=\hat{\mathbf{v}}\mathrm{e}^{\mathrm{i}kz}$ and $\tau=\gamma t$
we can rewrite the equation (\ref{eq:pert_eq_gen}) in the following
form,
\begin{equation}
  \frac{\mathrm{d}}{\mathrm{d}\tau}\left[\begin{array}{c}
      \hat{v}_{x}\\
      \hat{v}_{y}
    \end{array}\right]=\left(2\Ro^{-1}+1\right)\left[\begin{array}{c}
      \hat{v}_{y}\\
      -\hat{v}_{x}
    \end{array}\right]+\varepsilon\left[\begin{array}{c}
      \hat{v}_{y}\\
      \hat{v}_{x}
    \end{array}\right],\label{eq:pert_eq_components}
\end{equation}
\noindent where $\Ro=\gamma/\Omega$ is the Rossby number. The first
term on the right hand side of (\ref{eq:pert_eq_components}) corresponds
only to local rotation and the second one to the local rotation and
strain (in the following we will refer to the first term as the `pure
rotation term' and to the second one as the `straining term'). We
can easily solve the above equations (\ref{eq:pert_eq_components})
to obtain
\begin{equation}
  \hat{v}_{x}=\frac{1}{2}\left(\chi^{-1}\hat{v}_{y0}+\hat{v}_{x0}\right)\mathrm{e}^{\sigma\tau/\gamma}+\frac{1}{2}\left(-\chi^{-1}\hat{v}_{y0}+\hat{v}_{x0}\right)\mathrm{e}^{-\sigma\tau/\gamma},\label{eq:pert_sol_x}
\end{equation}
\begin{equation}
  \hat{v}_{y}=\frac{1}{2}\left(\hat{v}_{y0}+\chi\hat{v}_{x0}\right)\mathrm{e}^{\sigma\tau/\gamma}-\frac{1}{2}\left(-\hat{v}_{y0}+\chi\hat{v}_{x0}\right)\mathrm{e}^{-\sigma\tau/\gamma},\label{eq:pert_sol_y}
\end{equation}
\noindent where we have assumed that $\varepsilon^{2}>\left(1+2\Ro^{-1}\right)^{2}$
(thus the system is unstable) and
\begin{equation}
  \chi=\sqrt{\frac{\varepsilon-\left(1+2\Ro^{-1}\right)}{\varepsilon+\left(1+2\Ro^{-1}\right)}},\qquad\qquad\frac{\sigma}{\gamma}=\sqrt{\varepsilon^{2}-\left(1+2\Ro^{-1}\right)^{2}},\label{eq:chi_and_grate}
\end{equation}
\noindent (see (\ref{eq:grate_1})). The constants $\hat{v}_{x0}$
and $\hat{v}_{y0}$ are the initial values of the components
of the perturbation. It can easily be seen that in the case when no
strain is present, i.e. $\varepsilon=0$, the only term that remains
in the equation (\ref{eq:pert_eq_components}) corresponds to local
rotation and there is no possibility for instability (cf. the growth
rate in (\ref{eq:chi_and_grate})). If, on the other hand, the pure
rotation term vanishes, i.e. $\Ro=-2$ and only the straining term
remains, the destabilisation is maximal. The unstable mode, which
dominates the dynamics at large $\tau$ has a direction defined by
\begin{figure}
  \begin{center}
    \resizebox{.5\textwidth}{!}{\includegraphics{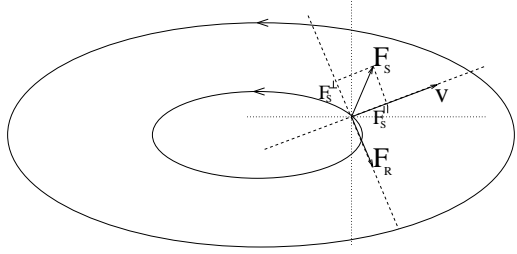}}
  \end{center}
  \caption[]{A sketch of the forces acting on fluid element
in the presence of horizontal perturbation.}
  \label{fig:meri_stab_mech}
\end{figure}
\begin{equation}
  \frac{\hat{v}_{y}^{(\rm{unstable})}}{\hat{v}_{x}^{(\rm{unstable})}}=\chi.\label{eq:unstable_direction}
\end{equation}
\noindent In general the straining term possesses not only the component
parallel to the perturbation but also the perpendicular one (see figure
2) and they are given by
\begin{equation}
  F_{S}^{\parallel}=\frac{2\varepsilon\hat{v}_{x}\hat{v}_{y}}{v},\label{eq:F_parallel}
\end{equation}
\begin{equation}
  F_{S}^{\perp}=\varepsilon\frac{\hat{v}_{x}^{2}-\hat{v}_{y}^{2}}{v},\label{eq:F_perp}
\end{equation}
\noindent where $v=\sqrt{\hat{v}_{x}^{2}+\hat{v}_{y}^{2}}$
and the magnitude of the pure rotation term, which is always perpendicular
to the velocity vector of the perturbation is
\begin{equation}
  F_{R}^{2}=\left(1+2\Ro^{-1}\right)^{2}v^{2}.\label{eq:F_rot}
\end{equation}
\noindent The parallel component $F_{S}^{\parallel}$ can either amplify
the perturbation or damp it depending on the sign of $\hat{v}_{x}\hat{v}_{y}$,
hence if the two perpendicular forces $F_{S}^{\perp}$ and $F_{R}$
can not be in balance for any $\hat{v}_{x}$ and $\hat{v}_{y}$,
the perturbation will not grow for all times, but it will oscillate.
The perpendicular forces can only be in balance if
\begin{equation}
  \frac{\varepsilon^{2}}{\left(1+2\Ro^{-1}\right)^{2}}=\left(\frac{v^{2}}{\hat{v}_{x}^{2}-\hat{v}_{y}^{2}}\right)^{2}>1,\label{eq:balance_perp}
\end{equation}
\noindent and then the perturbation grows unboundedly, amplified by
the parallel component $F_{S}^{\parallel}$ (cf. \fig{fig:meri_stab_mech}). 
The equation (\ref{eq:balance_perp}) is solved by any perturbation satisfying
(\ref{eq:unstable_direction}), which means that if the balance between
the perpendicular forces can occur for some $\hat{v}_{x}^{(\rm{unstable})}$
and $\hat{v}_{y}^{(\rm{unstable})}$, any random perturbation will
tend to the unstable mode defined by (\ref{eq:unstable_direction})
by rotating the perturbation vector towards the unstable direction
(and only one direction is stable, for which the perpendicular forces
are also in balance but the parallel force acts in opposite direction
to the direction of the perturbation velocity vector damping the perturbation,
i.e. for $\hat{v}_{y}^{(\rm{stable})}/\hat{v}_{x}^{(\rm{stable})}=-\chi$).
Furthermore, the equation (\ref{eq:balance_perp}) means that destabilisation
can only occur if $\varepsilon^{2}>\left(1+2\Ro^{-1}\right)^{2}$, which
is the instability requirement obtained in (\ref{eq:chi_and_grate}).

In the rotating shear flow limit, $\varepsilon=1$, which as shown earlier
corresponds locally to the centrifugal instability with horizontal
modes, the unstable direction is determined by $\chi=\sqrt{-1/(1+\Ro)}$.
Using the transformations from the previous section (\ref{eq:trans1})
we can obtain the form of the unstable mode in the non-rotating frame,
in the vicinity of $\mathbf{r}'=\mathbf{R}'$,
\begin{equation}
  \hat{v}_{r}^{\prime}\approx C\sqrt{\frac{-1}{1+\Ro}}\mathrm{e}^{\sigma t},\qquad\hat{v}_{\varphi}^{\prime}\approx C\mathrm{e}^{\sigma t},\label{eq:vr_and_vphi_nr}
\end{equation}
\noindent where $C$ is constant and the cylindrical coordinates where
used. The unstable direction corresponds exactly to the unstable direction
in the classical centrifugal instability, given by
\begin{equation}
  \frac{\hat{v}_{\varphi}^{\prime}}{\hat{v}_{r}^{\prime}}=-\frac{1}{\sigma}\left(2\Omega\left(r'\right)+r'\frac{\mathrm{d}\Omega}{\mathrm{d}r'}\right)=-\frac{2-s}{\sigma}\Omega\left(r'\right)=\sqrt{-\left(1+\Ro\right)},\label{eq:unstab_dir_centifug}
\end{equation}
\noindent where $\Omega(r')\sim1/r^{\prime s}$ and the expression (\ref{eq:grate_CI_local})
were used.

In the magnetic case the Lorentz force enters the balance, which depends
on the wavelength of the perturbations through the electric currents. Therefore the flows which
are stable in the non-magnetic case may become unstable in the presence
of magnetic field if the wavelength of the perturbation can be sufficiently adjusted. The Navier-Stokes and the induction equations for the horizontal rotational MEI take the form
\begin{equation}
  \frac{\partial\mathbf{v}}{\partial t}=-\left(\mathbf{v}\cdot\nabla\right)\mathbf{u}_0-2\mathbf{\Omega}\times\mathbf{v}+\mathrm{i}h\mathbf{b},\label{eq:pert_eq_mag_NS}
\end{equation}
\begin{equation}
  \frac{\partial\mathbf{b}}{\partial t}=\left(\mathbf{b}\cdot\nabla\right)\mathbf{u}_0+\mathrm{i}h\mathbf{v},\label{eq:pert_eq_mag_IND}
\end{equation}
For $\varepsilon=1$ they correspond to the MRI equations in Balbus \& Hawley (1991) with the radial wave vector (and thus the vertical components of the perturbations) set to zero (in their notation $k_R =0$). Of course such restriction makes the considerations of Balbus \& Hawley (1991) less general. However, it leads to the same stability criterion $\mathrm{d}\Omega/\mathrm{d}R >0$. 

As previously we can separate the total force on the right hand side of (\ref{eq:pert_eq_mag_NS}) into a component along the perturbation vector $\mathbf{v}$,
\begin{equation}
  F_{total}^{\parallel}=\frac{\mathrm{i}h}{v}\mathbf{v}\cdot\mathbf{b}+\frac{2\varepsilon v_x v_y}{v},\label{eq:F_tot_parallel}
\end{equation}
and a component perpendicular to $\mathbf{v}$
\begin{equation}
  F_{total}^{\perp}=\frac{\mathrm{i}h}{v}\left(v_x b_y-v_y b_x\right)+\frac{v_x^2-v_y^2}{v}-v-2\Ro^{-1}v.\label{eq:F_tot_perp}
\end{equation}
Next we find a general form of a $j$-th eigenvector of the matrix in (\ref{eq:HI_equations}) associated with the eigenvalue $\sigma_j$ ($j=1,2,3,4$),
\begin{subequations}\label{eq:eig_vect}
\begin{equation}
\hat{v}_{x}^{j} = -2\Ro^{-1}h\sigma_{j},\end{equation}
\begin{equation}
\hat{v}_{y}^{j} = h\left[\left(\varepsilon+1\right)\left(\varepsilon-1-2\Ro^{-1}\right)-h^{2}-\sigma_{j}^{2}\right],\end{equation}
\begin{equation}
\hat{b}_{x}^{j} = -i\left[\sigma_{j}^{2}\left(1+\varepsilon\right)+h^{2}\left(\varepsilon+1+2\Ro^{-1}\right)-\left(\varepsilon+1\right)\left(\varepsilon^{2}-\left(1+2\Ro^{-1}\right)^{2}\right)\right],\end{equation}
\begin{equation}
\hat{b}_{y}^{j} = -i\sigma_{j}\left[\varepsilon^{2}-\left(1+2\Ro^{-1}\right)^{2}-h^{2}-\sigma_{j}^{2}\right].\end{equation}
\end{subequations}
Since for any of the four eigenmodes $\mathbf{v}^{j}\cdot\mathbf{b}^{j}=p_c(\sigma_j)=0$, where $p_c(\cdot)$ is the characteristic polynomial of the matrix in (\ref{eq:HI_equations}) we get (for a single eigenmode)
\begin{equation}
F_{total}^{j\parallel}=-4\varepsilon \Ro^{-1}\left[\left(\varepsilon+1\right)\left(\varepsilon-1-2\Ro^{-1}\right)-\sigma_j^2-h^2 \right].\label{eq:F_parallel_gen_mode}\end{equation} 
We now concentrate on the unstable mode with index $1$. Inserting $\sigma_j=\sigma_1$ (cf. (\ref{eq:HIsigma})) into the above expression one can easily show that $F_{total}^{1\parallel}>0$ if $\sigma_1>0$, i.e. for $\varepsilon^2>\psi_{+}^2$. 

The calculation of the force component perpendicular to the perturbation velocity vector for the unstable mode $1$ is slightly more involved, but it is straightforward to show that
\begin{equation}
F_{total}^{1\perp}=0.\label{F_perp_unstab_mode}\end{equation}
Thus the mechanism of the instability is essentially the same as in the non-magnetic case, i.e. for the instability to be excited there must exist a possibility for a balance of all the rotational components of the forces acting on the perturbation, namely strain, the Coriolis and Lorentz forces, so that the perturbation for which such balance occurs can be amplified constantly in time by the strain. If such a balance cannot occur, the perturbations are amplified and damped in finite periods of time, thus they oscillate and do not grow. The unstable direction can be obtained from (\ref{eq:eig_vect}a,b). It can be seen, that in the magnetic case the unstable direction depends on the wavelength of the perturbation. This is no surprise, since it was already mentioned that since the Lorentz force depends on the wavelength through the currents, the balance of the rotational components depends on the choice of $k$. This explains why an elliptic vortex, which is stable in the non-magnetic case, may become unstable when the magnetic field is turned on with respect to perturbations for which the wavelengths
must satisfy a certain condition. For the case of rotating shear flow,
i.e. $\varepsilon=1$, a system with any negative Rossby number ($\Ro<0$)
is unstable with respect to horizontal perturbations with the wave
number $k$ satisfying (cf. (\ref{eq:grate_1})) $k^{2}<-4\Ro\Omega^{2}/v_{A}^{2}$. In a vertically bounded domain, such as an accretion disk, there is a lower bound on the allowed values of $k$, thus such conditions are, in principle, modified (cf. Bajer \& Mizerski 2011).

\section{Conclusions}

We have studied the properties and physical meaning of the horizontal MERI, i.e. the instability of the Euler flow with elliptical streamlines in the presence of external magnetic field and background rotation with respect to horizontal perturbations propagating along the field lines. The presence of anticyclonic background rotation is necessary for excitation of the horizontal instability and we have demonstrated that adding the magnetic field changes the range of values of the Rossby number at which the instability develops from $\Ro<-1$ in the non-magnetic case to $-4\gamma^2/(k^2 v_{A}^{2}) < \Ro < 0$ when the magnetic field is switched on.

The main purpose of this paper is the unification of the two, previously unrelated instability types. By the use of the local shearing box approximation the direct correspondence in the stability characteristics was shown between the well known short-wavelength MRI and the horizontal rotational MEI in the limit of rotating shear flow. The physical conditions under which the direct link between the two instabilities exists and the correspondence of the parameters between both instability types were explained in detail. The exponent $s$ defining the radial dependence of the angular velocity in the disc, $\Omega(r)\sim1/r^s$, is related to the local Rossby number in the box by $s=-2\Ro$ and the MRI requirement for instability, namely $s>0$ corresponds to the rotational MEI requirement, i.e. $\Ro<0$. Moreover, the dispersion relation for rotating shear flow with axial magnetic field for the horizontal modes was shown to be exactly the same as for the short-wavelength MRI and the well-known feature of the MRI in Keplerian discs, that the instability occurs for $0<kv_A/\Omega<\sqrt{3}$ with the maximal growth rate $\sigma=3\Omega/4$ achieved for $kv_A/\Omega=\sqrt{15}/4$ is also observed in rotating shear flow. The non-magnetic case of the horizontal instability was shown to have the same stability characteristics in the limit of rotating shear flow as the local centrifugal instability, with the Rayleigh's criterion corresponding to the non-magnetic horizontal instability requirement $\Ro<-1$.

\noindent Finally, we have also provided a physical interpretation of the horizontal instability, based on the balance between the strain generated by the basic elliptic flow and the Coriolis force. The unstable and stable directions of the perturbation velocity field as functions of $\Ro$ and $\varepsilon$ were found and the correspondence to the unstable direction of the perturbations for localised centrifugal instability was shown. It was also demonstrated that instability can only occur, if there is a possibility for the Coriolis force to be balanced by a component of the strain to eliminate the rotational tendency in the perturbation velocity, which, if present for all times, leads to oscillations. Such balance defines the unstable/stable direction and can occur only if $\varepsilon^{2}>\left(1+2\Ro^{-1}\right)^{2}$, which agrees with the results of the linear analysis. 

\noindent The full magnetic problem, for which the instability mechanism is essentially the same, was also discussed, highlighting the effect of the Lorentz force, which significantly affects the force balance, making the unstable direction dependent on the wavelength of the perturbation. The fact that the Lorentz force can be adjusted by the choice of the perturbation wavelength results in a possibility of achieving the balance of the rotational components of all three forces, which leads to continuous amplification of the perturbation, for negative Rossby numbers even outside the non-magnetic interval $2/(\varepsilon-1)<\Ro<-2/(\varepsilon+1)$. In particular in the rotating shear flow limit, $\varepsilon=1$ all systems with $\Ro<0$ are unstable, hence the presence of magnetic field leads to destabilisation of all elliptical vortices with $\Ro<-1$.
\\

WL gratefully acknowledges partial financial support of the US-American National Science Foundation (NSF) under grant no. AST10-09802.

\bibliographystyle{jfm}

\end{document}